\documentclass[a4,12pt]{article}
\newcommand{\be}{\begin{eqnarray}}
\newcommand{\ee}{\end{eqnarray}}
\makeatletter
\newcommand{\fslash}[2][0mu]{%
    \mathchoice
     {\fsl@sh\displaystyle{#1}{#2}}%
     {\fsl@sh\textstyle{#1}{#2}}%
     {\fsl@sh\scriptstyle{#1}{#2}}%
      {\fsl@sh\scriptscriptstyle{#1}{#2}}}
    \newcommand{\fsl@sh}[3]{%
    \m@th\ooalign{$\hfil#1\mkern#2/\hfil$\crcr$#1#3$}}
\def\lsim{\raise0.3ex\hbox{$<$\kern-0.75em\raise-1.1ex\hbox{$\sim$}}}
\def\gsim{\raise0.3ex\hbox{$>$\kern-0.75em\raise-1.1ex\hbox{$\sim$}}}
\makeatother
\usepackage{epsfig}
\usepackage{graphics}

\title{\bf Thermodynamics at Non-Zero Baryon \\
Number Density: A Comparison of 
Lattice and Hadron Resonance Gas \\
Model Calculations
   }
  \author{ \\ F. Karsch$^1$,
  K. Redlich$^{1,2}$, and A. Tawfik${^1}$
    \\ \\
    \small
 $^1$ Fakult\"at f\"ur Physik, Universit\"at Bielefeld,\\
\small Postfach 100 131, D-33501 Bielefeld, Germany\\
  \small     $^2$ Institute of Theoretical Physics University of Wroclaw,\\
\small   PL-50204 Wroclaw, Poland }
\begin{document}
\maketitle
{\vspace{-13.0cm}
\mbox{} \hfill BI-TP 2003/16\\
\vspace{12.5cm}
}

\centerline{Abstract} {\small We compare recent lattice studies of  
QCD thermodynamics at non-zero quark chemical potential
with the thermodynamics of a hadron resonance gas.
We argue that for $T\leq T_c$
the equation of state derived from Monte--Carlo simulations of two
flavour QCD at non-zero chemical potential can be well
described by a hadron resonance gas when using the same set of
approximations as used in current lattice calculations. 
We estimate the importance of truncation errors arising from the 
use of a Taylor expansion in terms of the quark chemical potential
and examine the influence of unphysically large quark masses
on the equation of state and the critical conditions for
deconfinement. }

\newpage
\section{Introduction}

While the thermodynamics of strongly interacting matter at
vanishing baryon number density or chemical potential has been
studied \cite{schladming} in lattice calculations for quite some
time, the   first investigations of the equation of state at
non-vanishing quark chemical potential ($\mu_q$) have started only
recently \cite{fodor,gavai,ejiri}. These studies of bulk thermodynamics
have been performed with different lattice actions and also have used
different methods (exact matrix inversion \cite{fodor} or Taylor
expansion \cite{gavai,ejiri}) to extend previous calculations performed
at $\mu_q = 0$ into the region of $\mu_q > 0$. Nonetheless they
led to qualitatively and even quantitatively similar results.

The basic pattern found for the temperature dependence of, e.g. the
pressure at finite density, follows closely that already seen at 
$\mu_q = 0$; the
pressure changes rapidly in a narrow temperature interval and
comes close to the Stefan-Boltzmann value of an ideal gas of
quarks and gluons at about twice the transition temperature.
Consequently, the density dependence of the equation of state in
the high temperature plasma phase has successfully been compared
\cite{quasi} with quasi-particle models that were  also used
 at $\mu_q = 0$.

In this work we concentrate on a discussion of the thermodynamics
of the hadronic phase of QCD in the regime of low baryon number density 
($\mu_q /T \lsim 1$) but high temperature ($T\sim T_c(\mu_q=0)$). 
In a recent paper we have shown \cite{recent} that the partition function 
of a hadron resonance gas  yields quite a satisfactory description of 
lattice results on bulk thermodynamic observables
in the low temperature, hadronic phase of QCD at $\mu_q = 0$.  We  will
extend here our previous study to finite chemical potential and
compare the predictions of the resonance gas model calculations
with lattice results obtained in 2-flavour QCD using a Taylor
expansion for small $\mu_q/T$ \cite{ejiri}. The reference system
for these calculations is  a previous analysis  \cite{karsch1} of
the temperature dependence of the pressure in 2-flavour QCD
performed at $\mu_q = 0$. Unlike the approach based on an exact
inversion of the fermion determinant \cite{fodor} the Taylor
expansion, obviously, has the disadvantage of being approximate.
There is, however, good reason to expect, that at least at high
temperature, the contribution of terms that are beyond ${\cal
O}((\mu_q/T)^4)$ order  is small. Moreover, as will become clear
from our discussion below, it turns out that the expansion
coefficients themselves provide useful information on the relevant
degrees of freedom controlling the density dependence of
thermodynamic quantities. We argue that baryons and their
resonances are these relevant degrees of freedom that govern
thermodynamics in the confined phase at finite density. We show
that for $T\leq T_c$ the  equation of state  at non-zero 
chemical potential which has been obtained in lattice calculations
can be well described by a baryonic  resonance gas when using the
same set of approximations as used in current lattice studies. We examine 
the importance of truncation effects in the Taylor expansion and discuss
the influence of unphysically large quark mass values on thermodynamic 
observables  and the critical conditions for deconfinement.

\section{Finite density QCD and Taylor Expansion}

\begin{figure}[htb]
\vskip -.2cm
\begin{minipage}[t]{49mm}
%\hskip -0.2cm\
\includegraphics[width=14.89pc,height=18.5pc,angle=180]{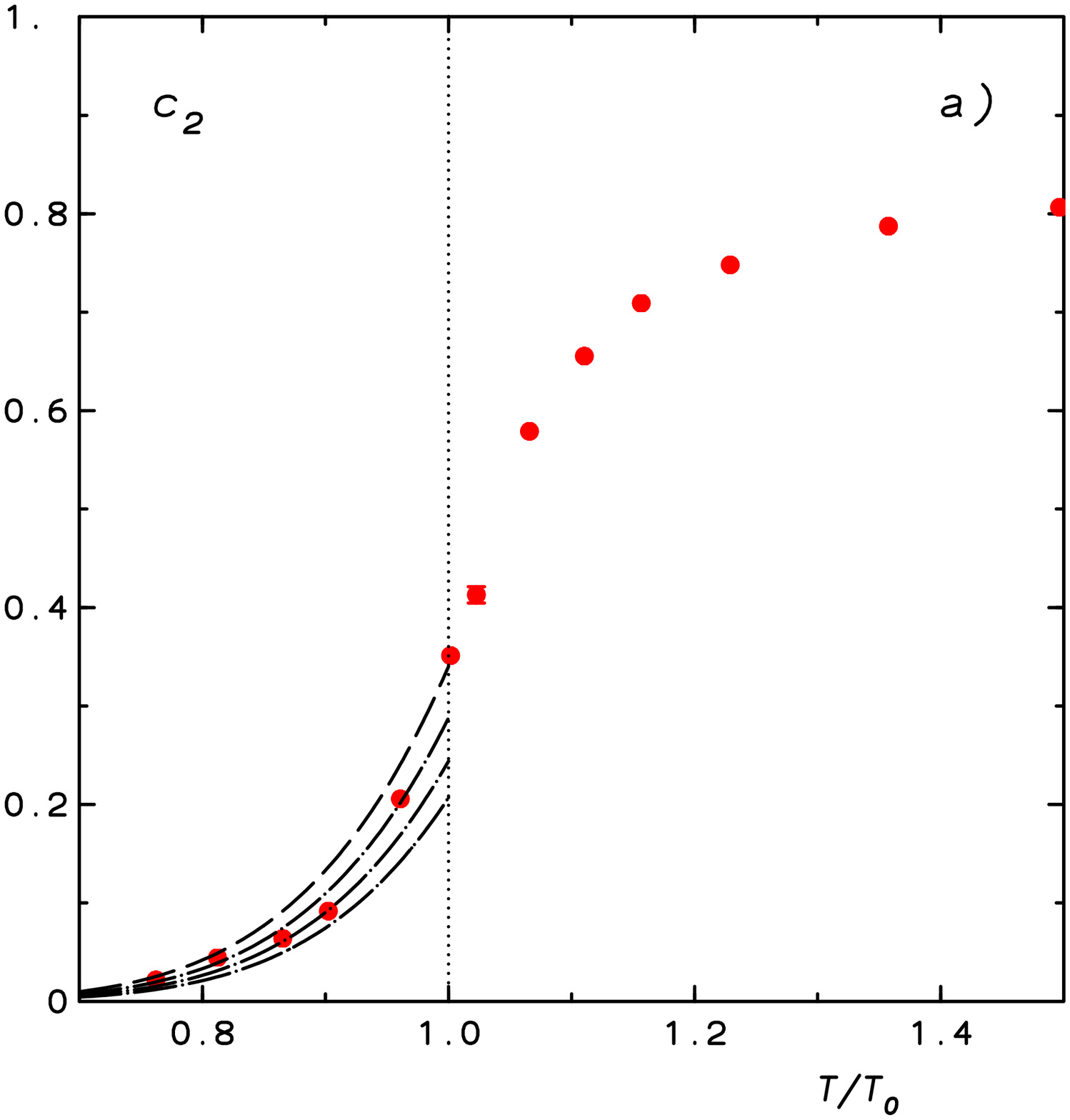}
\end{minipage}
\hskip 0.4cm
\begin{minipage}[t]{58mm}
%\hskip 0.5cm
\hskip 1.4cm
 \includegraphics[width=15.pc, height=18.3pc,angle=180]{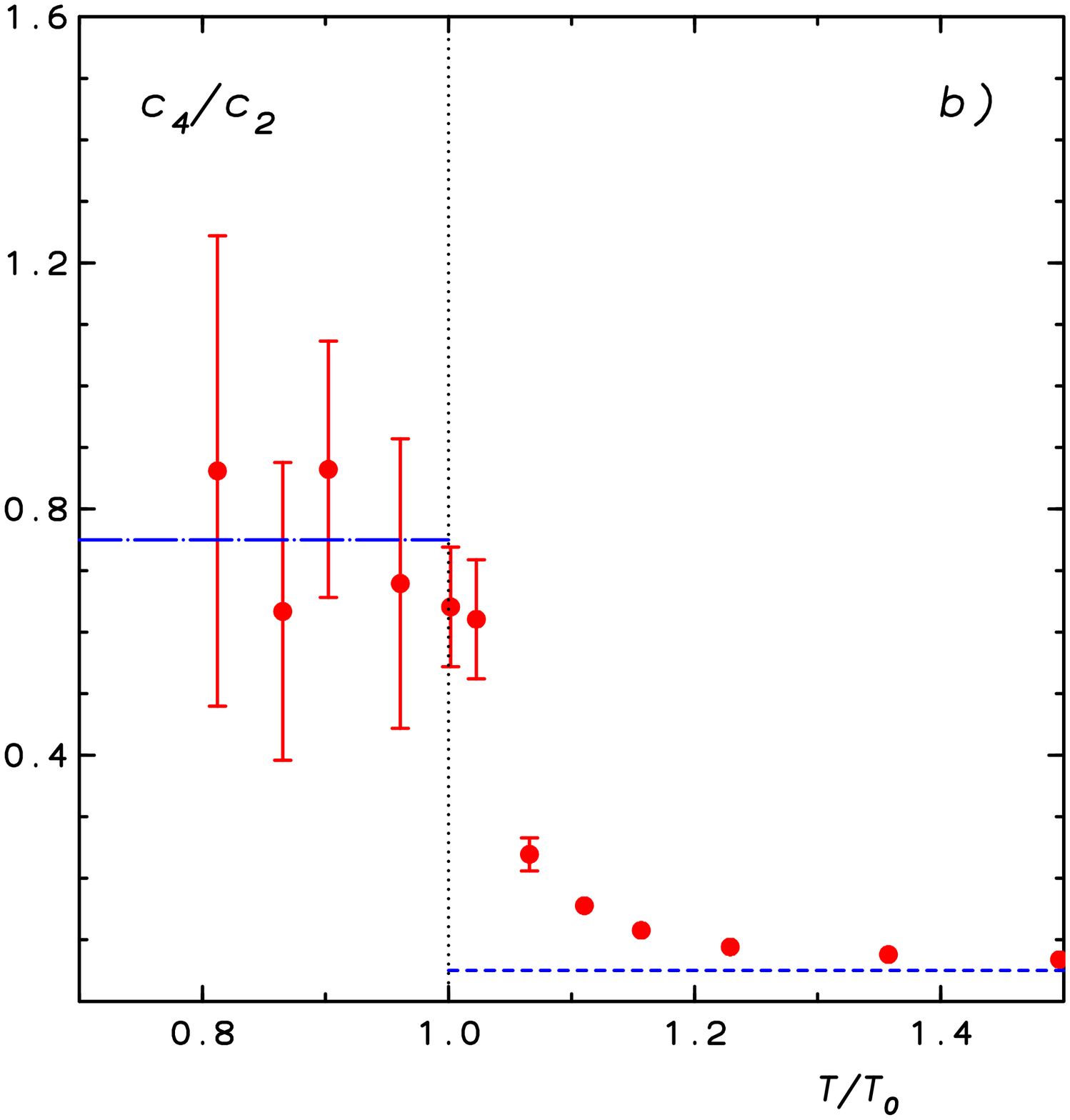}\\
\end{minipage}
\vskip -0.5cm \caption{ \label{fig:c2c4} 
Temperature dependence of the second order expansion coefficients $c_2$
and $c_4$ 
of the pressure in powers of $\mu_q/T$ obtained in 2-flavour QCD
\protect\cite{ejiri}. In Fig.~1a we show $c_2(T)$ and Fig.~1b shows
the ratio 
$c_4(T)/c_2(T)$. The temperature scale is given in units of the
transition temperature at $\mu_q=0$ which for the quark masses
used in the calculation of Ref.~\protect{\cite{karsch1}} is $T_0
\simeq 200$~MeV. For $T>T_0$ the dashed horizontal line shows the
massless Fermi gas value of $c_4/c_2$. The dashed-dotted curves in (a)
show results of a resonance gas model calculation for $A=0.9, 1.0, 1.1,
1.2$ (from top to bottom) as discussed in section 3. The dashed-dotted 
curve in (b) represents the resonance gas value $c_4/c_2=0.75$. }
\end{figure}

The basic quantity
that describes   thermodynamics at non-vanishing chemical
potential is the pressure. In the grand--canonical ensemble it is
obtained\footnote{Although the volume dependence of thermodynamic
quantities calculated on the lattice requires a careful analysis
and has not yet been performed for most thermodynamic studies with
dynamical quarks, we will for simplicity of notation suppress in
the following any volume dependence in our formulas.} from the
partition function as
\begin{equation} p(T,\mu_q) = \lim_{V \rightarrow
\infty} {T\over V} \ln Z(T,\mu_q, V) \quad . \label{pressure}
\end{equation}
For small values of $\mu_q / T$ the pressure may be expanded in a
power series,
\begin{equation}
{p(T,\mu_q) \over T^4} = \sum_{n=0}^{\infty} c_{2n}(T) \left( {\mu_q \over
T} \right)^{2n} \quad .
\label{series}
\end{equation}
In recent lattice calculations this series has been analyzed up to
${\cal O} ((\mu_q / T)^4)$ and 
in addition to the density dependent change of the pressure ($\Delta p$)
quantities like the quark number density
($n_q$) and quark number susceptibility ($\chi_q$) have been calculated. 
The latter are obtained from Eq.~\ref{series}
through appropriate derivatives with respect to the quark chemical
potential, 
\begin{eqnarray}
{\Delta p \over T^4} &=& {p(T,\mu_q) - p(T,0) \over T^4} \simeq
c_2(T) \left( {\mu_q \over T} \right)^2 +
c_4(T) \left( {\mu_q \over T} \right)^4 \quad , \nonumber \\
{n_q \over T^3} &=& {\partial\; p(T,\mu_q) \over \partial\; \mu_q}
\simeq 2\; c_2(T) \left( {\mu_q \over T} \right) +
4\; c_4(T) \left( {\mu_q \over T} \right)^3 \quad , \nonumber \\
{\chi_q \over T^2} &=& {\partial^2\; p(T,\mu_q) \over \partial\;
\mu_q^2} \simeq 2\; c_2(T) + 12\; c_4(T) \left( {\mu_q \over T}
\right)^2 \quad . \label{seriesO4}
\end{eqnarray}
In the asymptotically high temperature limit  this expansion
terminates, in fact, at the order given in Eq.~\ref{seriesO4}. The
expansion coefficients are then given by $c_2(\infty)= n_f/2$ and
$c_4(\infty)/c_2(\infty)= 1/2\pi^2$ respectively\footnote{ Also
at $O(g^2)$ the high temperature expansion terminates at
$O(({{\mu_q}\over T})^4)$. This, however, changes in the resumed
$O(g^3)$ contribution. The complete expansion up to $O(g^6\ln g)$
has recently been presented in \cite{v}.}.  In this limit the ratio
$c_4/c_2$ is small and the leading order term, consequently,
dominates in the Taylor expansion for a wide range of values for
$\mu_q/T$. The lattice results for the expansion coefficients, 
obtained in 2-flavour QCD are
shown in Fig.~\ref{fig:c2c4}. It can be seen, that  at $T\simeq
1.5\;T_0$ the numerical results for $c_2(T)$ still deviate by
about 20\% from the ideal gas value while the ratio $c_4/c_2$  is
already close to the corresponding result expected in the infinite
temperature limit.

\section{Resonance Gas and Boltzmann Approximation}

The analysis of experimental data on the production cross sections
of various hadrons in heavy ion collisions shows astonishingly
good agreement with corresponding thermal abundances calculated
in a hadronic resonance gas model at appropriately chosen
temperature and chemical potential \cite{redlich}. Our recent
analysis of the equation of state calculated on the lattice at $\mu_q = 0$
also has shown that a gas of non-interacting resonances can provide a good
description of the low temperature phase of QCD \cite{recent}. We want to
extent here our analysis to the case of non-vanishing chemical potential.

The partition function of a resonance gas can be specified through
the mass spectra for the mesonic and baryonic
sectors of QCD, respectively. In a non-interacting
resonance gas
the partition function reads,

\be
 \ln Z(T,\mu_B, V)=\sum_{i\; \in\; {\rm mesons}} \ln Z_{m_i}^B(T,V) +
\sum_{i\; \in\; {\rm baryons}} \ln Z_{m_i}^F(T,\mu_B, V) \quad ,
\label{eqq1} \ee where $Z_{m_i}^B~(Z_{m_i}^F)$ denote single
particle partition functions for bosons and fermions with mass
$m_i$ and $\mu_B=3\mu_q$ is the baryon chemical potential. 
Here the fermion partition function contains the
contribution from a particle and its anti-particle. The total
pressure of the resonance gas is then obtained from
Eq.~\ref{pressure} and builds up as a sum of contributions from
particles of mass $m_i$. The dependence of the pressure on the
chemical potential at a fixed temperature is thus entirely due to
the baryonic sector. The contribution, $p_m$, of baryons of mass
$m$ to the total pressure is given by
\begin{equation}
\left.\frac{p_m}{T}\right. = \frac{d}
{2\pi^2} \int_0^\infty \!\!\!dk\,k^2
\ln\left[(1+z\exp\{-\varepsilon(k)/T\})
(1+z^{-1}\exp\{-\varepsilon(k)/T\})\right]
\label{pressure_con}
\end{equation}
where  $z\equiv\exp\{\mu_B/T\}$ is the baryonic fugacity with
$\mu_B=3\mu_q$, $d$ is the spin--isospin degeneracy factor  and
$\varepsilon(k)=\sqrt{k^2+m^2}$ is the relativistic single
particle energy. The pressure  may be expressed in terms of a
fugacity expansion as
\begin{equation}
\frac{p_m}{T^4} = \frac{d}{\pi^2} \left( \frac{m}{T} \right)^2
\sum_{\ell=1}^{\infty} (-1)^{\ell+1}\,\ell^{-2}\, K_2(\ell m/T)\,
\cosh (\ell \mu_B/T)\quad ,
\label{bessel}
\end{equation}
where $K_2$ is the Bessel function.

In the hadronic phase of QCD the relevant temperatures will always
be smaller than the transition temperature to the plasma phase
determined in lattice calculations at $\mu_B=0$, {\it i.e.} we are
interested in the regime $T \lsim 200$~MeV \cite{karsch1}. As the
mass of the lightest baryon ($m_N$)  is about five times larger
than this value, the Bessel functions appearing in
Eq.~\ref{bessel} can always be approximated by the asymptotic form
valid for large arguments, {\it i.e.} $K_2(x) \simeq
\sqrt{\pi / 2x}\; \exp (-x)$.
This shows that higher order terms in Eq.~\ref{bessel} are
suppressed by factors $\exp(-\ell (m-\mu_B)/T)$.  As long as
$(m_N-\mu_B)\; \gsim\; T$  the contribution of baryons to the
resonance gas partition function is thus well approximated by the
leading term in Eq.~\ref{bessel}, which constitutes the Boltzmann approximation.
In this case each baryon/anti-baryon pair
contributes to the pressure with
\begin{equation}
\frac{p_m}{T^4} = \frac{d}{\pi^2} \left( \frac{m}{T} \right)^2
K_2(m/T)\, \cosh (\mu_B/T)\quad .\label{boltzmann}
\end{equation}
Thus,  the total baryonic contribution to the pressure of a
resonance gas reads
\begin{equation}
{p_B \over T^4} = F(T) \, \cosh (\mu_B/T)\quad ,
\label{totalp}
\end{equation}
where $F(T)$ is defined by
\begin{equation}
F(T)\equiv \sum_i {d_i\over {\pi^2}}\left( {m_i\over T} \right)^2
K_2( {m_i/ T} ) \quad ,\label{FT}
\end{equation}
and the sum is taken over all known baryons and their resonances.

Current lattice studies of the hadronic phase of QCD
with non-vanishing chemical potential concentrate on a temperature
regime $0.8\; T_0\; \lsim\; T\; \lsim\; T_0$ with quark chemical
potentials $\mu_q\; \lsim\; T$. The baryon chemical potential
$\mu_B = 3\mu_q$ thus stays considerably smaller than the nucleon
mass. Under these conditions the Boltzmann approximation will be
applicable and the dependence of the pressure on the baryon
chemical potential will just be mediated  through a multiplicative
factor as given  in Eq.~\ref{totalp}. We stress that this simple
relation is independent of details of the mass spectrum as long as
all fermions are sufficiently heavy. The factorization of the part
depending on the mass spectrum and that depending on the chemical
potential, however, also relies on the assumption that
interactions among the hadrons and resonances are negligible. The
validity of this assumption can be verified through a direct
comparison with lattice calculations for $\mu_q > 0$ and for
$T<T_0$.

\begin{figure}[htb]
\vskip -.2cm
\begin{center}
\includegraphics[width=15.pc, height=18.3pc,angle=180]{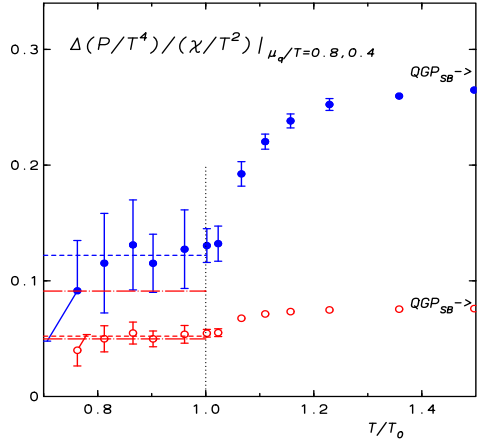}\\
\end{center}
%\end{minipage}
\vskip -0.7cm \caption{ \label{fig:ratio} The ratio of pressure
and quark number susceptibility versus temperature for fixed
values of $\mu_q / T$. The horizontal lines are the results of
hadron resonance gas model. The points are the lattice values from
Ref.~\protect\cite{ejiri}. While the dashed-dotted curves
represent the complete expression (Eq.~\protect{\ref{pchiratio}})
the dashed curves give the result of a Taylor expansion to the
same order as that used in the lattice calculations.}
\end{figure}

Following Eqs. \ref{boltzmann}--\ref{totalp} we can specify the
results for the change in pressure, the quark number density and
quark number susceptibility.  In order to compare the predictions
of  the resonance gas model  with lattice results one still needs  
to  perform the Taylor expansion up to the same  order as given in 
Eq.~\ref{seriesO4}. In the Boltzmann approximation we have,
\begin{eqnarray}
{\Delta p\over {T^4}} &=& F(T)[\cosh {{\mu_B}\over T}-1] \simeq
F(T) \left( \tilde{c}_2 \left( {\mu_q \over T}\right)^2
+ \tilde{c}_4 \left( {\mu_q \over T}\right)^4 \right)
\quad ,  \nonumber \\
{{n_{q}}\over {T^3}} &=& 3 F(T)\sinh {{\mu_B}\over T} \simeq
F(T) \left( 2 \tilde{c}_2 \left( {\mu_q \over T}\right)
+ 4 \tilde{c}_4 \left( {\mu_q \over T}\right)^3 \right)
\quad , \nonumber \\
{{\chi_{q}}\over {T^2}} &=& 9 F(T)\cosh {{\mu_B}\over T} \simeq
F(T) \left( 2 \tilde{c}_2 + 12 \tilde{c}_4 \left( {\mu_q \over T}\right)^2
\right)
\quad ,
\label{pnchi}
\end{eqnarray}
with $\tilde{c}_2 = 9/2$ and $\tilde{c}_4=27/8$. In the resonance
gas model the expansion coefficients introduced in
Eq.~\ref{seriesO4} are given by $c_{2n}= \tilde{c}_{2n} F(T)$. We
note that ratios of these quantities indeed are independent of the
resonance mass spectrum and only depend on the chemical potential.
For fixed $\mu_q/T$ we thus expect to find that any ratio of two
of the above quantities is temperature independent in the hadronic
phase. Using the same order of the Taylor expansion as used in the
lattice calculations such ratios only depend on
$\tilde{c}_4/\tilde{c}_2 = 3/4$, {\it i.e.} the resonance gas
model yields a temperature independent ratio $c_4/c_2$. As can be
seen in the right--hand part of Fig.~\ref{fig:c2c4} this is indeed
in good agreement with the lattice results. We note that  this
result is independent of details of the hadron mass spectrum. It
thus  should also   be insensitive to the change in the quark mass
used in the lattice calculation. In Fig.~\ref{fig:ratio} we show
the ratio $\Delta p/T^2\chi_q$ for two values of the chemical
potential. The good agreement between lattice calculations and the
hadronic gas results
merely reflects the  agreement found
already for the ratio $c_4/c_2$. In the resonance gas model we can,
however, provide also the complete result for this ratio,

%%%%%%%%%%%%%%%%%%%%%%%%%%%%%%%%%%

%%%%%%%%%%%%%%%%%%%%%%%%%%%%%%%%%%

\begin{equation}
{\Delta p \over T^2 \chi_q} = \frac{1}{9}
\left( 1 - \cosh^{-1}(3 \mu_q /T) \right) \quad .
\label{pchiratio}
\end{equation}
This is shown as a dashed--dotted  line in Fig.~\ref{fig:ratio}.

The  agreement of the ratio $c_4/c_2$  calculated in the resonance
gas model and on the lattice also implies that, for all values of
$\mu_q/T$, the temperature dependence of $\Delta p$ and its
derivatives like e.g. the quark number susceptibility is to a
large extent controlled by the same function, $c_2(T)= \frac{9}{2}
F(T)$.   To determine, however,   the function $F(T)$ we have to
specify the baryon mass spectrum explicitly. This is, in general,
known experimentally. However, to facilitate a direct comparison
with lattice calculations we have to take into account that the
spectrum is distorted due to the still quite large quark masses
used in calculations of the equation of state\footnote{The phase
transition temperature has been calculated at a large set of quark
mass values, including rather small values which lead to
''almost'' physical hadron masses. So far the equation of state,
however, has been studied with improved actions only for one set
of quark masses corresponding to a pion mass of about 770~MeV
\cite{karsch1}.}. In addition we have to take into account that
also the transition temperature is quark mass dependent. For the
quark mass value used in the numerical study   of the equation of
state \cite{ejiri,karsch1} the transition temperature has been
determined as $T_0 \simeq 200$~MeV.  While an extrapolation to the
chiral limit yields the transition temperature of about 170~MeV.
We use $T_0$ to express the hadron masses in units of the
temperature, $m/T \equiv (m/T_0)\cdot (T_0/T)$ with $T$ scaled in
units of $T_0$.% as given from the lattice calculations.

The distortion of the hadron mass spectrum due to unphysically large quark 
masses ($m_q$) can in general be deduced from lattice calculations at zero
temperature. A generic feature of such studies is that the deviation
from the physical mass value due to unphysically large values of the quark 
mass becomes smaller for heavier hadronic states (see eg. \cite{cppacs}).
Moreover, one finds \cite{qcdsf,schi} that the quark mass dependence is 
well parametrized through the relation, 
$(m_Ha)^2 = (m_Ha)^2_{phys} + b (m_\pi a)^2$,
where $(m_Ha)_{phys}$ denotes the physical mass value of a hadron
expressed in lattice units and $(m_Ha)$ is the value calculated on the
lattice for a certain value of the quark mass or equivalently a certain
value of the pion mass ($m_\pi^2 \sim m_q$). Until now, however, the masses 
of only a few baryonic states constructed from $(u, d)$-quarks have been 
studied in more detail on the lattice \cite{cppacs,qcdsf,schi}. This is 
obviously not sufficient to fix the function $F(T)$ in Eq.(\ref{boltzmann}) 
that requires the contributions from a large set of baryonic resonances.

The above quadratic parametrization of the quark mass dependence of
baryon masses shows at least for nucleon, delta and their parity partners  
only a weak dependence on the hadron mass.
We thus take this as a general ansatz for 
the parametrization of the dependence of baryon masses on the pion mass,

\begin{equation}
{{m (m_\pi)}\over m_H}\simeq 1+A\; {{m_\pi^2 }\over {m_H^2}} \quad ,
\label{par}
\end{equation}
where $m(m_\pi)$ is the distorted hadron mass at fixed $m_\pi$ and
$m_H$ is its corresponding physical value. This parametrization is
consistent with our previous analysis \cite{recent} where we have used
the MIT bag model to determine the $m_\pi$--dependence of hadron masses.

The lattice results \cite{ejiri} for QCD thermodynamics at finite
$\mu_q$ were obtained in 2-flavour QCD. An immediate consequence of
the restriction to only two quark flavour is that we have to suppress
the contribution from strange baryons to the resonance gas in the low 
temperature phase. Moreover, since the
lattice results were obtained with a quite large value of $m_q$,
corresponding to $m_\pi/\sqrt\sigma = 1.84\pm 0.04$, we 
account for modifications of the baryon mass spectrum using
Eq.~(\ref{par}). From the pion mass dependence of
the nucleon, delta and their parity partners \cite{qcdsf,schi}
we estimate $0.9\; \lsim\; A\; \lsim\; 1.2$ in Eq. (\ref{par}).
This range of values is also expected from the bag model study in Ref.
\cite{recent}.

\begin{figure}[htb]
\vskip -.2cm
\begin{center}
 \includegraphics[width=15.pc, height=18.3pc,angle=180]{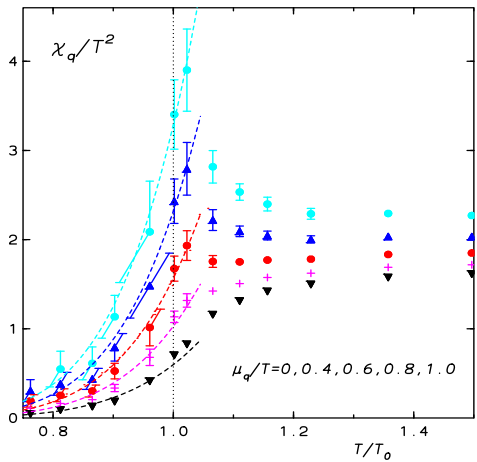}\\
\end{center}
\vskip -0.7cm \caption{ \label{fig:sus} Lattice results from
Ref.~\protect\cite{ejiri} for the quark number susceptibility in
2-flavour QCD calculated in next-to-leading order Taylor expansion
for different values of the quark chemical potential. The lines
are results from the resonance gas model using a distorted baryon
spectrum (Eq.~\ref{par} with $A=1$) and treated
within the same approximation as in the lattice study.}
\end{figure}

The above discussion fixes our parametrization of the baryonic
sector of the resonance gas model at unphysical values of the
quark mass as they are used in current lattice calculations. The
resulting temperature dependence of the expansion coefficient
$c_2(T)$ is shown in Fig.~\ref{fig:c2c4}a.
The corresponding result for the quark number susceptibility at
different values of the quark chemical potential is shown in
Fig.~\ref{fig:sus} for the choice $A=1.0$. The agreement of the resonance 
gas model and results obtained from lattice calculations
is indeed quite satisfactory. This indicates that the thermodynamics
of the confined phase of QCD at finite density is to large extent
governed by the baryonic resonances. As can be seen in Fig.~\ref{fig:c2c4}a 
the 15\% uncertainty in the parametrization of the baryonic mass
spectrum (Eq.~\ref{par}) results in $20 \%$  error on the values of
physical observables, {\it i.e.} $c_2(T)$, at $T=T_0$.
%%%%%%%%%%%%%%%%%%%%%%%%%%%%%%%%%%%%%%%%%%%%%%%%%%%%%%%

%%%%%%%%%%%%%%%%%%%%%%%%%%%%%%%%%%%%%%%%%%%%%%%%%%%%%%%

\section{Relaxing the lattice constraints}

\begin{figure}[htb]
\vskip -.2cm
\begin{minipage}[t]{49mm}
%\hskip -0.2cm\
\includegraphics[width=14.89pc,height=18.5pc,angle=180]{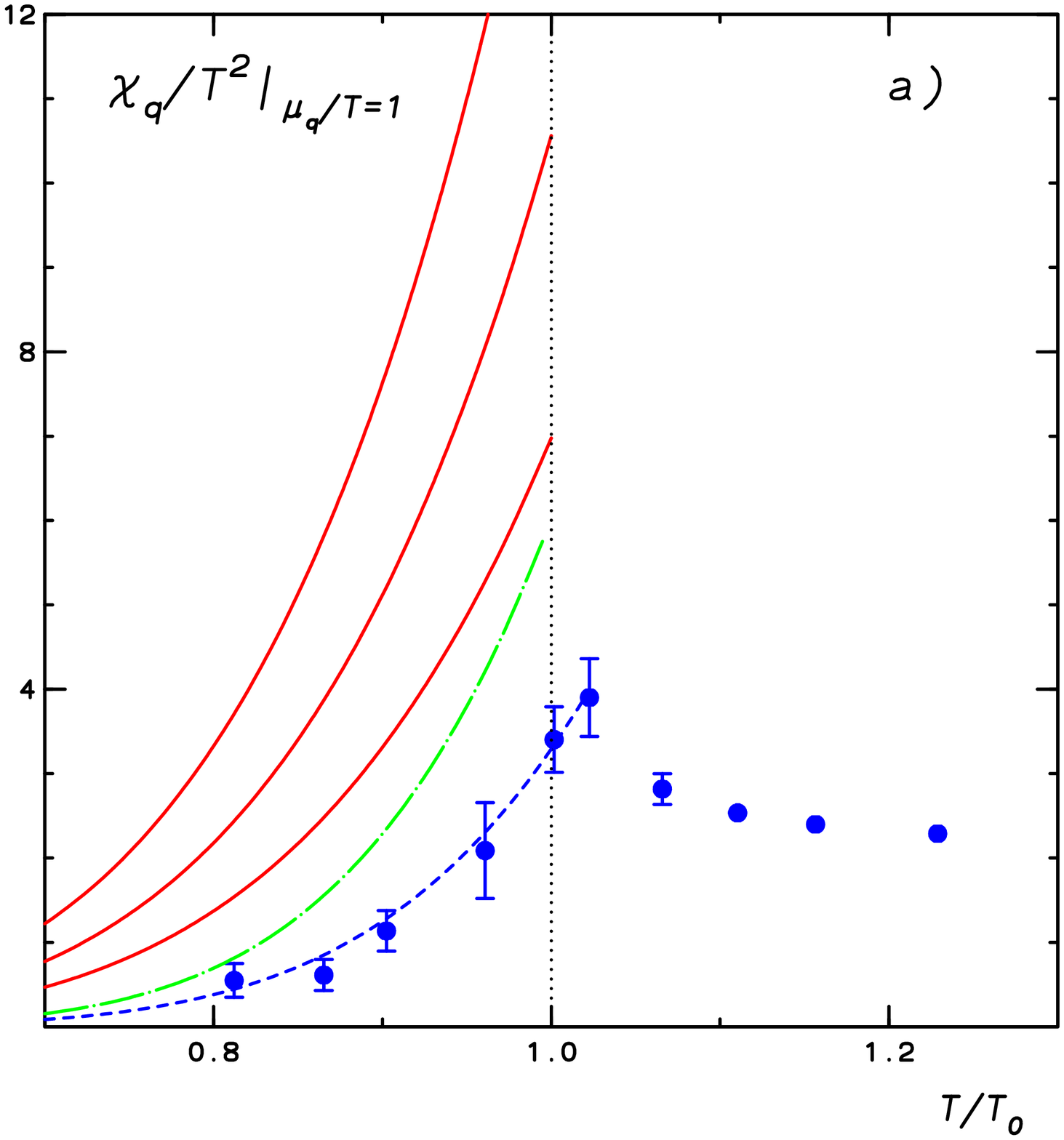}
\end{minipage}
\hskip 0.4cm
\begin{minipage}[t]{58mm}
%\hskip 0.5cm
\hskip 1.4cm
 \includegraphics[width=15.pc, height=18.3pc,angle=180]{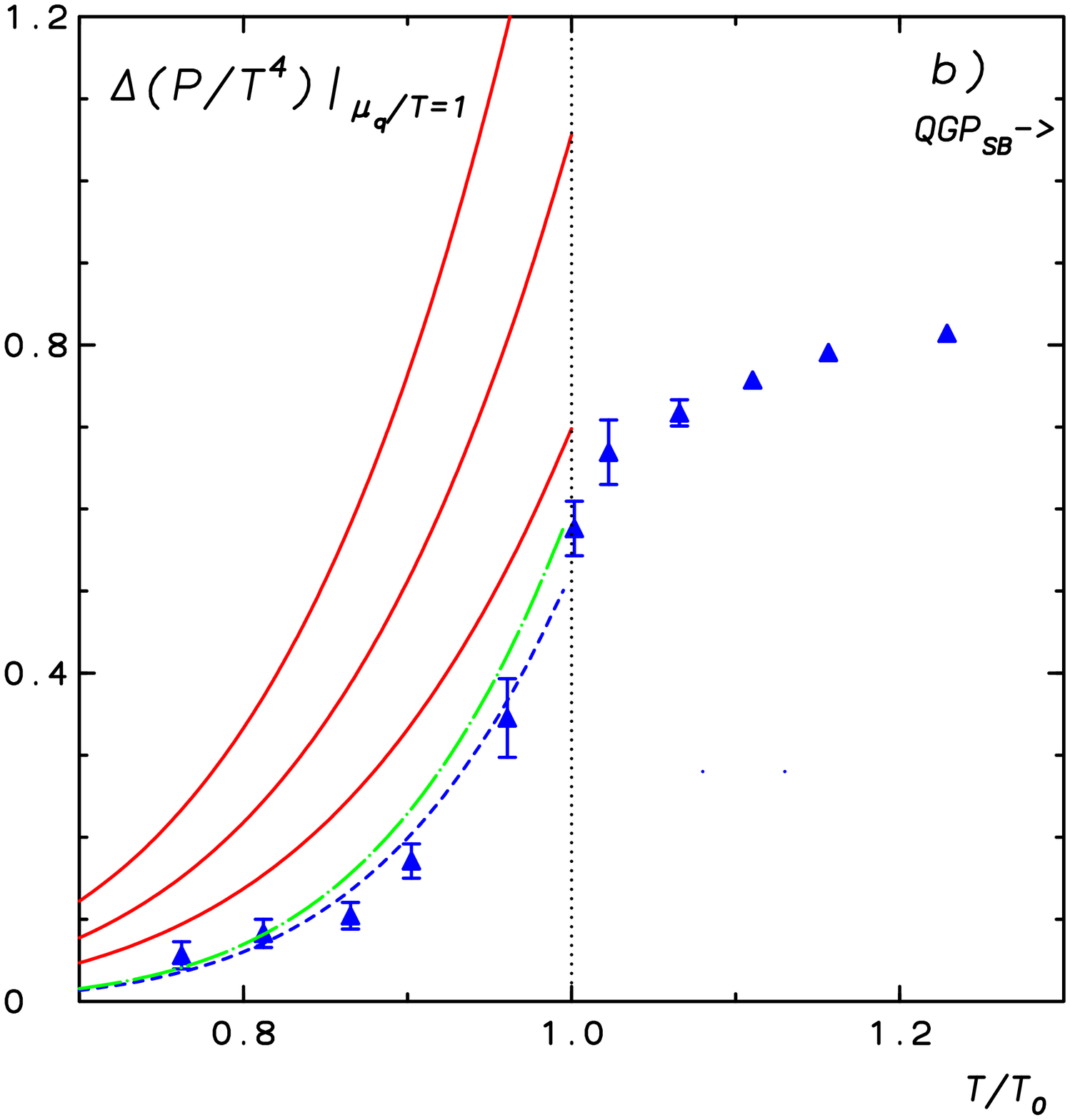}\\
\end{minipage}
%\vskip -0.5cm
\caption{ \label{fig:exp} Quark number susceptibility (a) and 
change in pressure (b)  for fixed quark chemical potential
$\mu_q/T=1$ as a function of $T/T_0$.  The points are lattice
results from Ref.~\protect{\cite{ejiri}} and lines are the
resonance gas model results. The dashed and dashed dotted lines are 
obtained with a
baryon mass spectrum appropriate for the unphysical quark masses
used in lattice calculations and with a Taylor expansion truncated at
${\cal O}((\mu_q/T)^4)$ and the full result, respectively. The
full lines are resonance gas model results obtained with
physical hadron masses, no expansion in $\mu_q/T$ and for three values
of the transition temperature, $T_0= 160$~MeV (lower), 170~MeV (middle)
and 180~MeV (upper), which cover the range of current lattice 
estimates for the chiral limit extrapolation of $T_c$ in 2-flavour QCD.}
\end{figure}

We want to discuss here in somewhat more detail what the resonance
gas model calculation suggests for the quark mass dependence of
current thermodynamic studies on the lattice and what the
influence of the truncation of the Taylor series expansion on the
behavior of thermodynamic quantities in the hadronic phase could
be. The latter clearly depends on the observable under consideration.
As anticipated also in Ref.~\cite{ejiri} the influence of a truncation 
of the Taylor expansion for the pressure at ${\cal O}((\mu_q /T)^4)$ becomes 
more severe in calculations of the quark number susceptibility 
as the expansion stops here already at ${\cal O}((\mu_q /T)^2)$.
The resonance gas model suggests that for $\mu_q / T = 1$ the truncated 
result for the pressure differs only by 15\% from the complete result.
These truncation errors rise to about 80\% in calculations of
$\chi_q/T^2$ at $\mu_q/T = 1$. The major part of this
truncation error could be removed by calculating the ${\cal
O}((\mu_q /T)^6)$ contribution to $\Delta p /T^4$. These
properties are seen in  Fig.~\ref{fig:exp} where the results of
the resonance gas calculation performed with the complete 
$(\mu_q/T)$-dependence and a Taylor expansion
truncated at $O((\mu_q/T)^4)$ are shown
%and complete $O((\mu_q/T)^\infty)$
%predictions of the resonance gas partition function is indicated
for pressure and quark number susceptibility.

%%%%%%%%%%%%%%%%%%%%%%%%%%%%%%%%%%%%%%%%%%%%%%%%%%%%%%%

Already in the discussion of the $\Delta p / T^2 \chi_q$ we 
indicated that within the Boltzmann approximation this and similar 
ratios are independent of the resonance mass spectrum and thus
also on the quark masses. Changes in the quark mass thus will 
influence calculations of the pressure and its derivatives in a
similar way. Replacing in the resonance gas calculation the modified
baryon spectrum by the experimentally known spectrum will increase
the value of the pressure as all baryons become lighter.
This effect is somewhat reduced as also the relevant temperature scale
is shifted to smaller values, {\it i.e.} the transition temperature 
will shift from $T_0 \simeq 200$~MeV to $T_0 \simeq 170$~MeV. As can
be seen in Fig.~\ref{fig:exp} this increases the pressure and its 
derivatives. The extrapolation to the physical case depends, however,
quite sensitively on the value for the critical temperature. We note
that the resonance gas model calculation favours a small critical
temperature as it seems to be unlikely that $\Delta p /T^4$ will 
exceed the corresponding ideal gas value at $T_0$.

Finally, we want to comment on the convergence radius of the Taylor
expansion in $(\mu_q / T)$. The resonance gas model, of course, does
not lead to critical behaviour. Consequently also the dependence on
$(\mu_q / T)$ is given by an analytic function. The resulting Taylor
expansion has an infinite convergence radius, which in terms of the
convergence criterium used in Ref.~\cite{ejiri} is reflected in the fact
that,
\begin{eqnarray}
\lim_{n\rightarrow \infty}\sqrt{c_{2n}/c_{2n+2}} &=& \infty \quad ,
\nonumber 
\end{eqnarray}
for all temperatures in the resonance gas model. In the case of QCD,
we expect, however, that the convergence radius is bounded from above
by the location of the phase boundary to the quark gluon plasma phase.
In particular, we expect that the ratios $c_{2n}/c_{2n+2}$ stay close
to unity for temperatures close to $T_0$. In fact, this also is the case
for the low order expansion coefficients in the resonance gas model,
{\it i.e.} $c_2/c_4 =4/3$. This, however, changes already at the 
next order, $c_4/c_6 = 10/3$. We thus expect that differences between
lattice results on the QCD thermodynamics and resonance gas model
calculations should show up at ${\cal O}((\mu_q/T)^6)$.

\section{Conclusion and Outlook}

We have shown that basic features of the density dependence of the
QCD equation of state in the hadronic phase observed in recent 
lattice studies can be
understood in terms of the thermodynamics of a baryonic resonance
gas. A quite robust result, independent of the detailed structure
of the hadron mass spectrum, is the dependence of thermodynamic
quantities on $\mu_q/T$ at fixed temperature. This indicates that
the thermodynamics at low temperatures is dominated by heavy
degrees of freedom which justify a Boltzmann approximation
for the partition function. Within the resonance gas model these
degrees of freedom are non-interacting which leads to simple
relations among different thermodynamic observables. In
particular, we argued that the ratio $\Delta p / T^2\chi_q$ is
independent of temperature at fixed $\mu_q /T$. This is in
agreement with current lattice calculations. It is, however, clear
that this has to change, when the system undergoes a true phase
transition at some temperature $T_c(\mu_q) < T_0$ for a
sufficiently large value of $\mu_q /T$. In particular, we expect
that the quark number susceptibility will diverge at the chiral
critical point, i.e. at the second order endpoint of a line of first
order phase transitions \cite{Hatta}, which should lead to a dip
in the ratio $\Delta p / T^2\chi_q$ at $T_c(\mu_q)$.

Finally we note that the relevance of the resonance gas model
predictions for the description of the low temperature phase of QCD can
also be tested in the imaginary chemical potential approach \cite{philipsen}
which also is used in lattice  calculations. The resonance gas model, for
instance, suggests that the change in pressure is quite well described
by a simple trigonometric function, $\cos (3\mu_q /T)$.

\section*{Acknowledgments}
\addcontentsline{toc}{section}{Acknowledgements}

We  acknowledge  stimulating discussions with S. Ejiri and   C.
Schmidt. K.R. also acknowledges the support of the Alexander von
Humboldt Foundation (AvH) and the Polish Committee for Scientific Research
(KBN). This work has partly been supported by the DFG under grant 
FOR 339/2-1 and the GSI collaboration grant BI-KAR.

\end{document}